\documentclass[onecolumn,11pt,draftclsnofoot]{IEEEtran}
\usepackage{graphicx}
\graphicspath{ {images/} }
\usepackage{mathtools}
\usepackage{amsmath}
\usepackage{amsmath}
\usepackage{algorithm}
\usepackage[noend]{algpseudocode}
\usepackage{color,soul}
\usepackage{algorithm,algpseudocode}
\makeatletter
\newcommand{\algmargin}{\the\ALG@thistlm}
\makeatother
\newlength{\whilewidth}
\algdef{SE}[parWHILE]{parWhile}{EndparWhile}[1]
  {\parbox[t]{\dimexpr\linewidth-\algmargin}{%
     \hangindent\whilewidth\strut\algorithmicwhile\ #1\ \algorithmicdo\strut}}{\algorithmicend\ \algorithmicwhile}%
\algnewcommand{\parState}[1]{\State%
  \parbox[t]{\dimexpr\linewidth-\algmargin}{\strut #1\strut}}

\ifCLASSINFOpdf
\else
\fi
\newtheorem{theorem}{Theorem}

\makeatletter
\def \BState{\State \hskip - \ALG@thistlm}
\makeatother

\begin{document}
%
\title{Energy Efficient Distributed Worst Case Robust Power Allocation in Massive MIMO}
%
%
%

\author{
       Saeed~Sadeghi~Vilni
\thanks{S. Sadeghi was with the Department
of Electrical and Computer Engineering, Tarbiat Modares,Tehran,
 Iran  (e-mail: saeed.sadeghi@modares.ac.ir).}

}

\maketitle

\begin{abstract}
This letter proposes an energy efficient distributed worst case robust power allocation in massive multiple input multiple output (MIMO) system. We assume a bounded channel state information (CSI) error and all channel lie in some bounded uncertainty region. The problem is formulated as max-min one with infinite constraint. At first, we solve inner problem with triangle and Cauchy-Schwarz inequality, then by fractional programming and successive convex approximation (SCA) technique problem transfers to a convex optimization. Finally closed form transmit power is obtained with distribution way. Simulation results demonstrate proposed algorithm convergence and validate robust power allocation. Also appropriate number of transmit antenna to have maximum energy efficiency in simulation result is shown.
\end{abstract}
\IEEEpeerreviewmaketitle
\section{Introduction}
%
%
%
%
\IEEEPARstart{M}{assive} MIMO is one of main technology that candidate for next generation. Massive { MIMO} enhance spectral and energy efficiency regards to accessing CSI for efficient beamforming and appropriate number of transmit antenna to get high energy efficiency. Energy efficiency defined as spectral efficiency to power consumption ratio \cite{IEEEhowto:survey}. By increasing the number of transmit antenna spectral efficiency is increased while circuit power consumption is also increased. So if circuit power consumption don't considered in total power consumption, optimal number of transmit antenna is infinity \cite{circuit}.
To estimate channel, users in any cell send pilot to own base station ({BS}) then {BS} estimate users channel. In practice, the number of orthogonal pilot sequences are limited so users use pilots that are non-orthogonal related to users in the other cells. Thus precision of channel estimated, due to inter cell interference for users which utilizing non-orthogonal pilots, is decreased. This issue is named pilot contamination \cite{IEEEhowto:ch}.

In \cite{bojson}-\cite{pcup} energy efficient power allocation in a massive MIMO system for different scenarios are studied, where authors in \cite{bojson}-\cite{coner} consider pilot contamination but do not robust design. Total power consumption in \cite{MRTee}-\cite{pcup} is not determined properly. Circuit power consumption do not considered in \cite{MRTee} and assuming constant in \cite{pcup} where circuit power consumption is a function of number of transmit antenna.

Worst case robust approach in resource allocation has been considered in \cite{kpair}, \cite{eecogni}, \cite{mimopower} in different contexts. Author in \cite{kpair} proposed a robust transceiver design for the K-pair quasi-static MIMO interference channel with fairness considerations. they did their design as an optimization problem to maximize the worst-case SINR among all users. In \cite{eecogni}, investigated the robust energy efficiency maximization in underlay cognitive radio networks with bounded errors in all channels, and adopted the worst-case optimization approach to ensure primary users’ QoS requirement. Author in \cite{mimopower}, studied robust resource allocation schemes for MIMO-wireless power communication networks, where multiple users harvest energy from a dedicated power station in order to be able to transmit their information signals to an information receiving station.

In this letter energy efficiency maximization problem with circuit power consumption and pilot contamination considered which transmit power is designed robustly. We formulate optimization problem to maximize the minimum energy efficiency related to channel estimation error bound and under constraint on power transmit and meet QoS. After finding minimum of energy efficiency respect to uncertainty region of estimation error we find optimal transmit power to maximize energy efficiency, and finally, an algorithm to find optimal transmit power is proposed. To find optimal number of transmit antenna, maximum energy efficiency regards to optimization problem for different transmit antenna is plotted. Simulation result validate effectiveness and convergence of algorithm.

{\it Notation:} The superscript $H$ stand for conjugate transpose. $I_{K}$ is the $K\times K$ identity matrix and $0_L$ is the $L \times 1$ all zero vector. $\| . \|$  represents Euclidean norm and $(x)^+=max(0,x)$. ${\bf n}\sim \mathcal{CN} (0,\bf C)$ means probability density function of zero mean complex Gaussian vector with covariance matrix {\bf C}.

\begin {figure}[h] \label{fig:p1}
\centering
\includegraphics[scale=0.2]{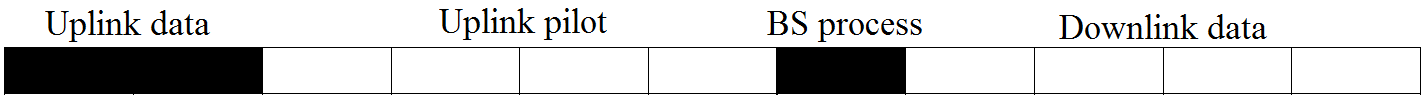}
\center
\caption {TDD protocol for massive MIMO in a coherent interval}
\end {figure}\section{System Model}

We consider the downlink of a multi-cell network with \textit{L} cells, where any cells utilize from massive {MIMO}.
 BS have\textit{ M} antenna with a linear configuration, and any {BS} serves \textit{K} single antenna users where randomly located in each cell. All {BS}s and users use from a time frequency resource and operate in time division duplex (TDD) mode as shown in figure 1.

\subsection{Channel Model }
\ In the part of uplink pilot in coherent interval, any user sends $\tau$ pilot symbols with the power $q$
\ and then each {BS}s estimate the channels of its users. The pilot sequence of cell \textit{l}  represented by a $K \times \tau$ matrix $S_l$. Pilot sequence of a cell users are orthogonal then $S_l S_l^H= I_K$. Due to pilot reuse   multiple of $S_i$ for different cell is not always zero. The received signal at {BS} $i$ is represented by an $M \times \tau$ matrix $Y_i$ as\\
\begin{equation}
Y_i = \sqrt{q}H_i S + Z_i 
\end{equation}
Where $S=[S_1;...;S_L] \in C^{   ( K \times L ) \times \tau}$ and $H_i \in C^{M \times ( K \times L ) }$ is the channel matrix between all the users and $i$th BS, whose $k$th column of $H_i$ is $h_{ilk}$ represent the gains of the channels from user $k$ in cell $l$ to BS $i$ and $h_{ilk}=\sqrt{\beta_{ilk}}g_{ilk}$; where $\beta_{ilk}$ is large scale fading involve path loss and shadowing and $g_{ilk}$ is small scale fading with $\mathcal{CN}(0,1)$ distribution. $Z_i$ is an additive noise matrix with independent identically distributed and complex Gaussian random variable with zero mean and unit variance entries.

By using a linear filter, estimated channel is $\hat H_i=\frac{1}{\sqrt{q}} Y_i S^H$ and $h_{ilk} =\sqrt{\beta_{ilk}}\hat g_{ilk} $ where distribution of $\hat g_{ilk} $ is $\mathcal{CN}(0_M,(\sum_{l=1}^{L}\beta_{ilk} + 1)I_M)$. It is obvious that estimated channel is not adopt to real channel $h_{ilk}$ which lead to estimation error is denoted by $\Delta $, therefore real channel can be written as follows
\begin{equation}
h_{ilk} =  \hat h_{ilk} + \Delta
\end{equation}
We assume that the
actual channel $ h_{ilk}$ lies on the neighborhood of estimated channel $ \hat h_{ilk}$
that is known to the transmitter. We consider that $ h_{ilk}$
is in the uncertainty region $R$ with radius $\sqrt a$ that define as the following ellipsoid
\begin{equation}
R = \{ \Delta : \parallel \Delta \parallel ^2 < a\}
\end{equation}
\subsection{Power Consumption}
The network power consumption is include transmit power, circuit power of transmitters and users. Circuit Power consumption in BS has two parts, first part is constant power consumption $P^{fix}$, this part involves site cooling, control signaling, backhaul, local oscillator, channel estimation and processors \cite {circuit}. Second part involves required power for any antenna to run which shown with $P^{pa}$. Also we consider transceiver required power of of each user $P^{pu}$. Thus the circuit power consumption of each cell $P^c_l$ expressed as
\begin{equation}
P_l^c = P^{fix} + M P^{pa} + KP^{pu}
\end{equation}
\subsection{Energy Efficiency}
By the estimated channel and use of maximum ratio transmitter beamformer, beamforming vector for \textit {m}th user in the \textit {j}th cell  that be expressed as
\begin{equation}
w_{jm} = \frac{\hat h_{jjm}^H}{\parallel \hat h_{jjm} \parallel}
\end{equation}
the received signal at $m$th user in the cell \textit {j} given by
\begin{equation}
\begin{multlined}
y_{jm} = \sqrt{p_{jm}} w_{jm} h_{jjm} s_{jm} \\
 + \sum_{l=1,\neq j}^{L}\sum_{k=1,\neq m}^{K}\sqrt{p_{lk}} w_{lk} h_{ljm} s_{lk} + n_{jm}
\end{multlined}
\end{equation}
where $p_{lk}$ and $s_{lk}$ represent transmit power and data symbol for user \textit{k} in the cell $l$ respectively the $n_{jm}$ is noise at the $m$th user in the $j$th cell and assume noise power is $N_0$.  Received signal to interference plus noise ratio ({SINR}) of user $m$ in cell $j$ is obtained by
\begin{equation}
\gamma_{jm} = \frac{p_{jm} \| w_{jm} h_{jjm} \|^2}{\sum_{l=1}^{L}\sum_{k=1,\neq m}^{K} {p_{lk}} \|w_{lk} h_{ljm} \|^2 + N_0}
\end{equation}
then we can express the data rate for this user as
\begin{equation}
r_{jm} = B\log_{2}({1+\Gamma\gamma_{jm}})
\end{equation}
where $B$ is channel bandwidth and $\Gamma=\frac{-2}{3\ln (5e)}$ is SINR gap between Shannon channel capacity and practical situation, where $e$ is target bit error rate \cite{IEEEhowto:sgap}. Thus energy efficiency of the whole network expressed as
\begin{equation}
\eta = \frac{ \sum_{l=1}^{L}\sum_{k=1}^{K}r_{lk}}{\sum_{l=1}^{L}\sum_{k=1}^{K}p_{lk}+ \sum_{l=1}^{L}P_l^c}
\end{equation}
\subsection{Optimization Problem}
Based on worst case optimization, the energy efficiency maximization problem expressed as
\begin{equation}
\begin{array}{rlllll}
\displaystyle {\max_{\bf P}{  \min_{\Delta}} }& \multicolumn{1}{l}{\eta} \\
\textrm{s.t.} &  C1: \sum_{k=1}^{K}p_{lk}\leq P^{max}\\
&\displaystyle  C2: r_{lk}\geq R^{min}\\
& \displaystyle C3:\| \Delta \|^2 \leq a
\end{array}
\end{equation}
The goal of optimization problem is to find transmit power $\bf{P}$ $= [P_1,...,P_L]$ where $P_l=[p_{l1},...,p_{lK}]$, which optimize the worst energy efficiency for errors are in the uncertainty region. The optimization constraints are $C_1$ that shows maximum transmit power for each cell, $C_2$ shows minimum data rate requirements for any user and $C_3$ shows uncertainty region.
\section{Solution}
To solve the max-min problem, first inner minimization problem then the outer maximization are solved respectively.
\subsection{Worst Case SINR}
SINR is a fractional function of $\Delta$, thus to find minimum of SINR over uncertainty region we find minimum of the numerator and maximum of the denominator in uncertainty region. First we consider triangle inequality as follows
\begin{equation}
\begin{array}{rrr}
\| w_{jm}(\hat h_{jjm}+\Delta)\| ^2 \geq \| w_{jm}\hat h_{jjm}\| ^2 - \| w_{jm}\Delta\| ^2 \\
\| w_{jm}(\hat h_{jjm}+\Delta)\| ^2 \leq \| w_{jm}\hat h_{jjm}\| ^2 + \| w_{jm}\Delta\| ^2 
\end{array}
\end{equation}
respect to Cauchy-Schwarz inequality a lower bound of the SINR can be computed as follows

\begin{equation}\label{slower}
\gamma^{lb}_{jm} = \frac{p_{jm} ( \| w_{jm}\hat h_{jjm}\| ^2 - \| w_{jm}\| ^2 \| \Delta \| ^2)^+}{\sum_{l=1}^{L}\sum_{k=1,\neq m}^{K} {p_{lk}} ( \| w_{lk}\hat h_{ljm}\| ^2 + \| \Delta \| ^2\| w_{lk}\| ^2)^+ + N_0}
\end{equation}

 and with $\| \Delta \|^2 \leq a$ we have
\begin{equation}\label{CSI}
\begin{array}{rrr}
 \| w_{jm}\| ^2 \| \Delta \| ^2\leq a\| w_{jm}\| ^2
\end{array}
\end{equation}
According to inequalitys given in (\ref{slower}) and (\ref{CSI}) worst case SINR of user $m$ in cell $j$, $\gamma ^{*}_{jm}$ is computed as
\begin{equation}
\gamma^{*}_{jm} = \frac{p_{jm} ( \| w_{jm}\hat h_{jjm}\| ^2 - a\| w_{jm}\| ^2)^+}{\sum_{l=1}^{L}\sum_{k=1,\neq m}^{K} {p_{lk}} ( \| w_{lk}\hat h_{ljm}\| ^2 + a\| w_{lk}\| ^2)^+ + N_0}
\end{equation}
Now with worst case SINR we obtain worst case data rate $r^{*}_{jm}$ as follows
\begin{equation}
r^{*}_{jm} = B\log_{2}({1+\Gamma\gamma^{*}_{jm}})
\end{equation}
Then worst case energy efficiency over uncertainty region obtained as
\begin{equation}
\eta^{*} = \frac{ \sum_{l=1}^{L}\sum_{k=1}^{K}r^{*}_{lk}}{\sum_{l=1}^{L}\sum_{k=1}^{K}p_{lk}+ \sum_{l=1}^{L}P_l^c}=\frac{T({\bf{P}})}{E({\bf{P}})}
\end{equation}
which lead to Finally optimization problem given bellow
\begin{equation} \label {eq:16}
\begin{array}{rlllll}
\displaystyle {\max_{\bf{P}} }& \multicolumn{1}{l}{\eta^{*}} \\
\textrm{s.t.} &  C1: \sum_{k=1}^{K}p_{lk}\leq P^{max}\\
&\displaystyle  C2: r^{*}_{lk}\geq R^{min}
\end{array}
\end{equation}
\subsection{Problem Reformulation}
Optimization problem is a fractional problem, thus we use fractional programming method to solve it. We assume the answer of (\ref{eq:16}) is as power transmit ${\bf P}^*$ and maximum energy efficiency $\eta ^{op}$. Now we introduce following theorem based on Dinkelbach algorithm \cite{IEEEhowto:fractional}:
\begin{theorem}
The maximum energy efficiency $\eta ^{op}$ is achieved in (16) \textit{if and only if} $\displaystyle \max_{\bf P}$ $T({\bf{P}})-\eta ^{op}E({\bf{P}})=T({\bf{P}}^*)-\eta ^{op}E({\bf{P}}^*)=0$ for $T({\bf{P}})\geq0$ and $E({\bf{P}}) > 0$.
\end{theorem}
Thus problem (\ref{eq:16}) changed to following optimization problem
\begin{equation} \label {eq:17}
\begin{array}{rlllll}
\displaystyle {\max_{\bf{P}} }& \multicolumn{1}{l}{T({\bf{P}})-\eta^{*} E({\bf{P}})} \\
\textrm{s.t.} &  C1: \sum_{k=1}^{K}p_{lk}\leq P^{max}\\
&\displaystyle  C2: r^{*}_{lk}\geq R^{min}
\end{array}
\end{equation}
Now to solve problem (\ref{eq:16}) we should solve iteratively problem (\ref{eq:17}). For this a primary value of energy efficiency is considered to solve problem (\ref{eq:17}) then  $T({\bf{P}})-\eta^{*} E({\bf{P}})$ is computed, if it goes near to zero, $\eta^{*}$ is the optimal energy efficiency, else $\eta^{*}$ is computed respect to transmit power which obtained from solving (18), and do this iteratively until $T({\bf{P}})-\eta^{*} E({\bf{P}})$ goes very close to zero.

Our objective function is non convex. For transforming this problem to convex optimization one, successive convex approximation method is used \cite{IEEEhowto:scale}. In this method  following lower bound is assumed
\begin{equation}
\log(1+\Gamma\gamma)\geq \alpha \log(\Gamma\gamma) + \beta
\end{equation}
and considering $\bf P$ as $\hat {\bf P} = \log{\bf P}$. By utilizing SCA, optimization problem (\ref{eq:17}) transfers to a convex optimization problem. So final optimization problem expressed as
\begin{equation} \label {eq:19}
\begin{array}{rlllll}
\displaystyle {\max_{\hat {\bf{P}}} }& \multicolumn{1}{l}{T({\hat {\bf{P}}})-\eta^{*} E({\hat {\bf{P}}})} \\
\textrm{s.t.} &  C1: \sum_{k=1}^{K}e^{\hat{p}_{lk}}\leq P^{max}\\
&\displaystyle  C2: r^{*}_{lk}\geq R^{min}
\end{array}
\end{equation}
For solving problem (\ref{eq:17}) problem (\ref{eq:19}) is solved iteratively that use lower bound of $r_{lk}^{*}=\alpha_{lk} \log(\Gamma\gamma_{lk}^{*}) + \beta_{lk}$ then update $\alpha_{lk}$ and $\beta_{lk}$ in any iteration until power transmit converges. With assuming ${ \bf \hat P}^0$ as optimal transmit power value in latest iteration, optimal SINR $\gamma_{lk}^{0}$ is computed then update  $\alpha_{lk}$ and $\beta_{lk}$ as follows
\begin{equation}
\begin{array}{ll}
\alpha_{lk} = \frac{\gamma_{lk}^{0}}{1+\Gamma\gamma_{lk}^{0}}\\
\beta_{lk} = \log(1 + \Gamma\gamma_{lk}^{0}) -  \frac{\Gamma\gamma_{lk}^{0}}{1+\Gamma\gamma_{lk}^{0}} \log(\Gamma\gamma_{lk}^{0})
\end{array}
\end{equation}
When power transmit converge replace ${\bf \hat P}^0$ as optimal power transmit power.
\subsection{Optimal Power Allocation}
The Lagrangian function of (\ref{eq:19}) is obtained as
\begin{equation}
\begin{array}{lll}
\mathcal{L}(\hat{ \bf P},\lambda_{lk} , \mu_{l}) \\ = \sum_{l=1}^{L}\sum_{k=1}^{K}r_{lk}^{*}(\hat p_{lk})-\eta^{*}(\sum_{l=1}^{L}\sum_{k=1}^{K}e^{\hat p_{lk}} \\ + \sum_{l=1}^{L}P^{c}_{l}) +  \sum_{l=1}^{L}\sum_{k=1}^{K}\lambda_{lk}(r_{lk}^{*}(\hat p_{lk})-R^{min}) \\ -  \sum_{l=1}^{L}\mu_{l}(\sum_{k=1}^{K}e^{\hat p_{lk}}-P^{max})
\end{array}
\end{equation}
Where $\lambda_{lk}$ and $\mu_{l}$ are Lagrange multipliers corresponding to the two constraints.  Based on the Karush-Kuhn-Tucker ({KKT}) conditions optimal transmit power for user $m$ in cell $j$ the following condition must be satisfied\cite{boid}
\begin{equation}
\frac{\partial \mathcal{L}(\hat{ \bf P},\lambda_{lk} , \mu_{l})}{\partial {\hat p_{jm}}}=0
\end{equation}
Then the optimal transmit power for user $m$ in cell $j$ is obtained as follows
\begin{equation} \label{opt.P}
p_{jm} = (\frac{(\lambda_{jm}+1)\frac{B\alpha_{jm}}{\ln2}}{\frac{B}{\ln2}\sum_{l=1}^{L}\sum_{k=1,k\neq m}^{K}\alpha_{lk} (\lambda_{lk}+1)\frac{z_{lk}}{I_{lk}}-(\eta^{*} +\mu_j)})^+
\end{equation}
Where
\begin{equation}
z_{lk}= \|w_{jm}\hat h_{jlk}\|^2 + a\|w_{jm}\|^2
\end{equation}
\begin{equation} \label{Interference}
\begin{array}{lll}
I_{lk} = \sum_{n=1,n\neq l}^{L} \sum_{u=1,u\neq k}^{K} e^{\hat p_{nu}}(\|w_{nu}\hat h_{nlk}\|^2 + \|w_{nu}\| ^2) \\ +\|n_{lk}\| ^2
\end{array}
\end{equation}
$I_{lk}$ is estimation of Interference plus noise on user $k$ in cell $l$ that users compute and feed back to its {BS} and {BS}s share this value together.
from Lagrange function and use subgradient method the Lagrangian multipliers can be updated according to
\begin{equation} \label{eq:26}
\lambda_{lk}(t+1) =(\lambda_{lk}(t)-\zeta_0 (r_{lk}^{*}-R^{min}))^+ 
\end{equation}
\begin{equation}\label{eq:27}
\mu_{l}(t+1) =(\mu_{l}(t)-\beta_0 (P^{max}-\sum_{k = 1}^{K}p_{lk}))^+ 
\end{equation}
where $\zeta_{0}$ and $\beta_{0}$ are step size and $t$ is iteration index. finally, the algorithm of Energy Efficient Distributed Worst Case Robust Power Allocation presented in the Algorithm 1.


\section{Sum-rate maximization }
Maximum sum-rate under condition of problem (\ref{eq:16}) can be computed by setting the denominator of the energy efficiency equal to 1. The optimal transmit power for user $m$ in cell $j$, which maximize the network sum-rate, is obtained as follows
\begin{equation} \label{opt.P-rate}
p^{sr}_{jm} = (\frac{(\lambda_{jm}+1)\frac{B\alpha_{jm}}{\ln2}}{\frac{B}{\ln2}\sum_{l=1}^{L}\sum_{k=1,k\neq m}^{K}\alpha_{lk} (\lambda_{lk}+1)\frac{z_{lk}}{I_{lk}}-(\mu_j)})^+
\end{equation}
All the parameters are obtained as obtained for optimal transmit power for energy efficient power allocation. The algorithm of power allocation to maximize the sum-rate is done as power allocation in energy efficient power allocation in algorithm 1 with considering one iteration of Dinckelbach algorithm, because our problem is gone to a non-fractional problem.
\begin{algorithm} 
\caption{Energy Efficient Distributed Worst Case Robust Power Allocation}\label{euclid}
\begin{algorithmic}[1]
\State Initialize convergence tolerance $\epsilon$ and initialize arbitrary $\eta(t_1)=0$, ${con} = 0$ and $t_1=1$
\Repeat
\State initialize with a feasible ${\bf \hat P(t_2)}$, Set $\alpha_{lk}(t_2)=1$,
\parState{$\beta_{lk}(t_2)=0$ and  $t_{2}=1$}
\Repeat
\State  initialize arbitrary $\lambda_{lk}$, $\mu_{l}$, $\alpha_0$ and $\beta_0$
\Repeat
\parState{Compute $I_{lk}$ by user with (\ref{Interference}) and feed back to its {\bf BS} }
\parState{{\bf BS} $l$ compute summation of its users $I_{lk}$ and share with another {\bf BS}s}
\State Update $\lambda_{lk}$ and $\mu_{l}$ according to (\ref{eq:26}) and (\ref{eq:26})
\parState{Compute the optimal power transmit $\bf \hat P(t_2+1)$ according to (\ref{opt.P})}
\Until {Convergence of $\lambda_{lk}$ and $\mu_{l}$}
\parState{Compute $\gamma_{lk}^{*}(\bf \hat P(t_2+1))$ and update $\alpha_{lk}(t_2+1)$ and $\beta_{lk}(t_2+1)$}

\Until {convergence of $\bf \hat P$ }
\State Compute $\eta^*$ with optimal $\bf \hat P(t_2+1)$
\If {$T({\hat {\bf{P(t_2+1)}}})-\eta E({\hat {\bf{P(t_2+1)}}})\leq \epsilon $}
\State Set $con = 1$
\State {\bf Return} $\eta^*$ as optimal energy efficiency
\Else
\State Set $con = 0$
\State $t_1 = t_1 + 1$
\State {\bf Return} $\eta^*$ as initial $\eta(t_1)$ in iteration
\EndIf
\Until {${con} = 1$}
\end{algorithmic}
\end{algorithm}

\section{Simulation Result}
In this section we evaluate the proposed robust power allocation via simulation. First we show convergence of Algorithm {\bf 1}, then compare robust and non-robust design and finally represent cost of robustness. We consider a multi-cell cellular network with $L=3$ cells and a {BS} in its center. The {BS}s locate at the coordinate of $(0,0)$, $(0,1000)$ and $(0,2000)$. We consider large scale fading as $\beta_{ilk}=\phi(\frac{d_0}{d_{ilk}})^4$ where $\log(\phi)$ has a normal distribution with 0dB mean and 8dB variance, $d_{ilk}$ represent distance from user $k$ in cell $l$ to BS $i$. The radius of each cell is 500 meter. In each cell there are $K=5$ single antenna users that uniformly distributed in each cell and minimum distance of each user to its { BS} is $d_0=50$. We consider noise power $N_0=$-174dBm/Hz, bandwidth $B=180$KHz, $q = 10$w, $R^{min}=250$Kbps,$P^{max}=$1w, $P^{fix}=20$w, $P^{pa}=0.1$w, $e = 10^{-3}$ and $P^{pu}=0.01$w.
\begin {figure}[h]
\centering
\includegraphics[scale=0.4]{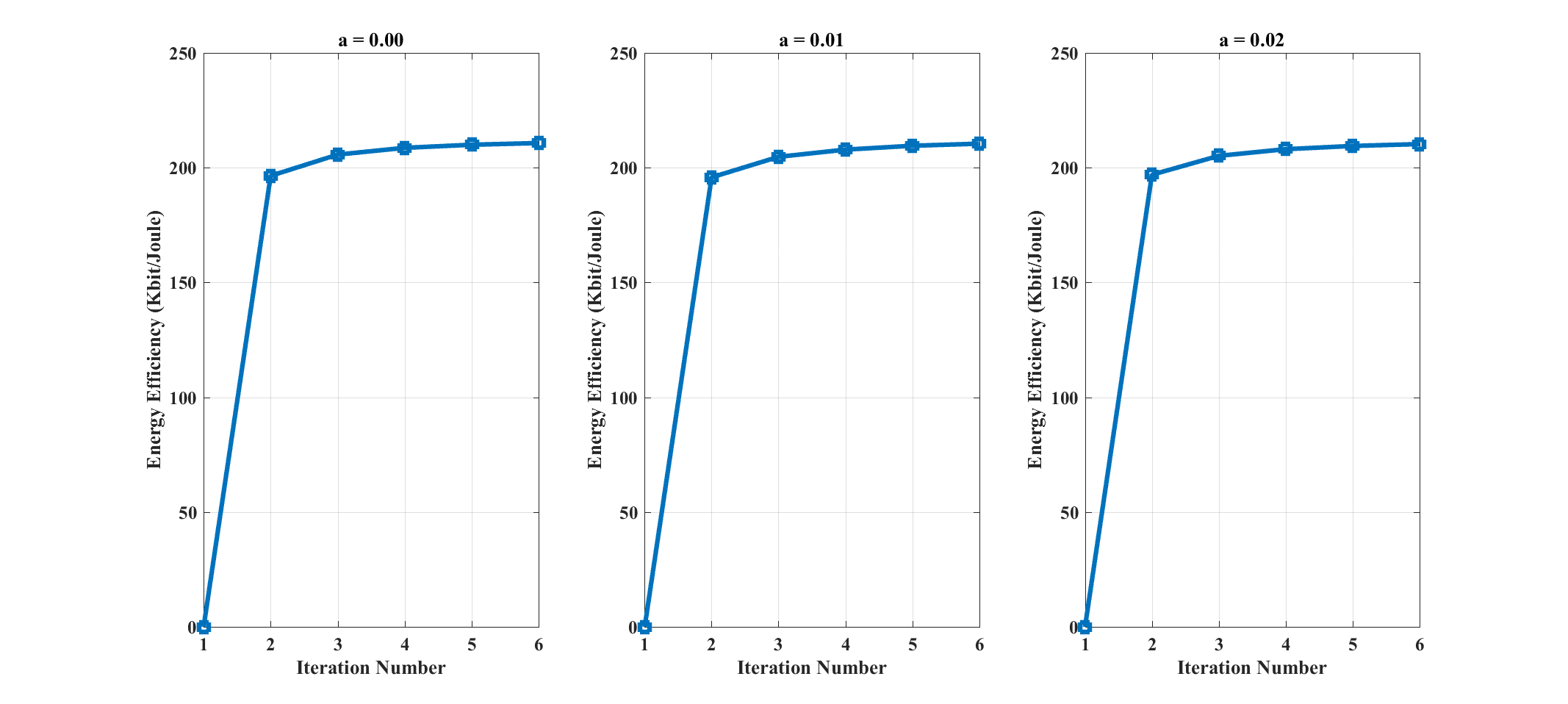}
\center
\caption {Convergence of Algorithm 1 for different uncertainty region with 100 antennas}
\label{fig:p1}
\end {figure}

\subsection{Convergence}
In fig.2 energy efficiency under Algorithm {\bf 1} over iterations respect to $M=100$ and three uncertainty region $a=0,0.01,0.02$ is shown. The energy efficiency converge to a fixed value after three iterations. Also uncertainty region does not affect on convergence iteration number.
\begin {figure}[h]
\centering
\includegraphics[scale=0.4]{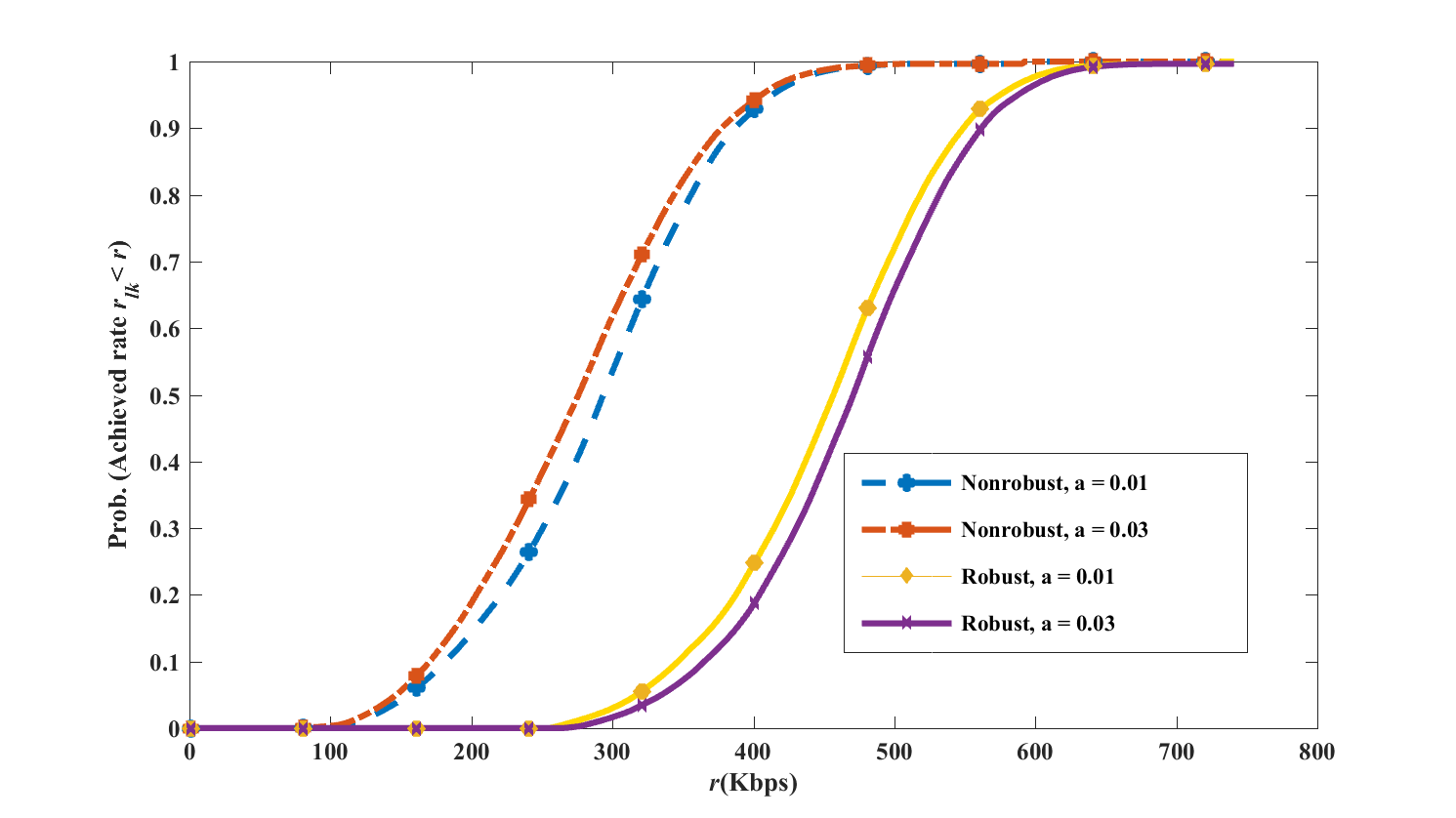}
\center
\caption {Performance of robust design with $R^{min}=215Kbps$}
\label{fig:p1}
\end {figure}
\subsection{Robustness Performance}
In fig.3 cumulative distribution function ({CDF}) of users data rate for robust and non-robust design while $R^{min}=14Kbps$ is shown. It obvious that in robust design size of uncertainty region have conversely relationship with probability of error, and vice versa for non-robust design.

\begin {figure}[h]
\centering
\includegraphics[scale=0.4]{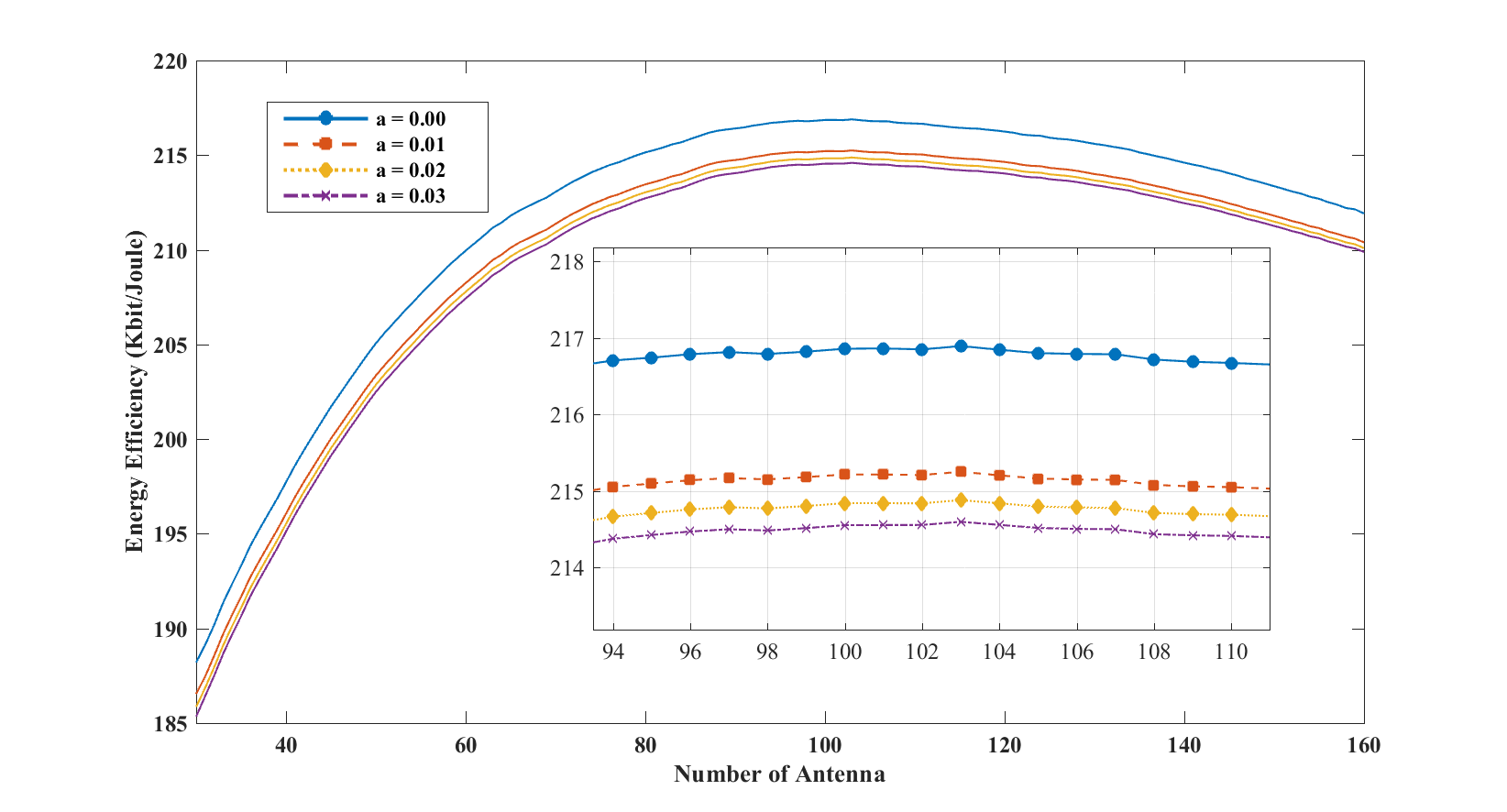}
\center
\caption {Energy efficiency over number of antenna for four different uncertainty region $a=0.00,0.01,0.02,0.03$ }
\label{fig:p1}
\end {figure}

\subsection{Cost of Robustness}
In fig.\ref{fig:p1} energy efficiency under Algorithm 1 over number of transmitter antenna is evaluated. It can be shown that by increasing radius of uncertainty region energy efficiency decreases. this decrease is cost of robustness. Also energy efficiency's curve first increases for dominance spectral efficiency increasing to circuit power consumption increasing where after $M = 102$ circuit power consumption increasing dominance spectral efficiency increasing, so appropriate number of transmit antenna obtaned equal to 102.

\section{Conclusion}
In this paper, we investigate energy efficient robust power allocation in a cellular network with massive MIMO BS. Based worst case approach we modeled channel then formulate our max-min energy efficiency problem. Max-min problem is solved in two step, first minimizing objective function on uncertainty region then maximizing on transmit power in a distributed way. Finally, a distributed robust power allocation to maximize energy efficiency proposed. Simulation result verify convergence of presented algorithm and worst case robust design performance. Also, simulation show, by increasing uncertainty region, energy efficiency is decreased and any number of transmit antenna is not appropriate

\end{document}